\documentstyle[epsf,a4]{article}  
\begin{document}                
\newcommand{\manual}{rm}        
\newcommand\bs{\char '134 }     

\newcommand{\simlt}{\stackrel{<}{{}_\sim}}
\newcommand{\simgt}{\stackrel{>}{{}_\sim}}
\newcommand{\MeV}{\;\mathrm{MeV}}
\newcommand{\TeV}{\;\mathrm{TeV}}
\newcommand{\GeV}{\;\mathrm{GeV}}
\newcommand{\eV}{\;\mathrm{eV}}
\newcommand{\cm}{\;\mathrm{cm}}
\newcommand{\s}{\;\mathrm{s}}
\newcommand{\sr}{\;\mathrm{sr}}
\newcommand{\lab}{\mathrm{lab}}
\newcommand{\sol}{{\mathrm{sol}}}
\newcommand{\atm}{{\mathrm{atm}}}
\newcommand{\ts}{\textstyle}
\newcommand{\ol}{\overline}
\newcommand{\be}{\begin{equation}}
\newcommand{\ee}{\end{equation}}
\newcommand{\ba}{\begin{eqnarray}}
\newcommand{\ea}{\end{eqnarray}}
\newcommand{\nn}{\nonumber}
\newcommand{\nm}{{\nu_\mu}}
\newcommand{\pp}{$\overline{p}(p)-p\;\;$}
\renewcommand{\floatpagefraction}{1.}
\renewcommand{\topfraction}{1.}
\renewcommand{\bottomfraction}{1.}
\renewcommand{\textfraction}{0.}               
\renewcommand{\thefootnote}{F\arabic{footnote}}
\title{Absolute values of three neutrino masses from atmospheric mixing
and an ansatz for the mixing-matrix elements}
\author{Saul Barshay and Georg Kreyerhoff\\III. Physikalisches Institut\\
RWTH Aachen\\D-52056 Aachen\\Germany}
\maketitle
\begin{abstract}                
Using data from atmospheric neutrino mixing, and a simple functional
form for mixing angles, the absolute values of three neutrino
masses are calculated: $m_3\cong 5.37\times 10^{-2}\eV$, $m_2=
(1.37-1.94)\times 10^{-2}\eV$, $m_1 = (1.03-1.46)\times 10^{-2}\eV$. The quantities
relevant for solar neutrino mixing are calculated: $(m_2^2-m_1^2)
= (0.8-1.63)\times 10^{-4}\eV^2$, with non-maximal mixing
$\tan^2\theta_\sol \cong 0.56$. The analysis gives a suggestion of a
dynamical origin for the empirical, large CP-violating phase
associated with an intrinsically, very small mixing angle in
the quark sector.
\end{abstract}
 
With the experimental determination \cite{ref1,ref2,ref3} of two
differences between squared masses for neutrinos, and nearly maximal
mixing \cite{ref1,ref4,ref5} for atmospheric neutrinos, the matter
of of the absolute value of the neutrino masses is of importance for
many physical processes \cite{ref6}. The purpose of this note is
to use the atmospheric measurements \cite{ref1} $\Delta m^2_\atm \cong
2.5\times 10^{-3}\eV^2$, $\tan^2\theta_\atm \cong 1$, together
with a simple ansatz for the functional form of the main elements
of the lepton mixing matrix which is motivated by certain analogies
to the quark sector, to determine three absolute values for neutrino
masses: $m_3\cong 5.37\times 10^{-2}\eV$, $m_2 =
(1.37-1.94)\times 10^{-2}\eV$, $m_1 = (1.03- 1.46)\times 10^{-2}\eV$.
These masses are hierarchal, although not as markedly so as the masses
of the three, charge (-1/3) quarks.
A further essential result following from our calculations is that the
second sizable mixing angle, responsible for the disappearence of
solar, electron neutrinos is not maximal (i.~e.~$45^\circ$), 
rather $\tan^2\theta_\sol \cong 0.56$. A comparison of the pattern
of lepton mixing to that in the quark sector shows that in both
cases, there appears to be a single, intrinsically, very small mixing
angle ($< 10^{-1}$ of either of the other two mixing angles). In
the quark mixing matrix, this very small angle is associated with
the Kobayashi-Maskawa \cite{ref7}, CP-violating phase $\delta_{13}$,
which as presently determined, may well be large \cite{ref8} ($\simgt 45^\circ$).
This can suggest that there may exist some kind of milli-weak interaction
which itself grossly violates CP invariance. This interaction dynamically
gives rise to a sizable CP-violating phase in the quark CKM matrix.
This matrix then phenomenologically describes the ``corrected'' weak interactions
with a third, very small mixing angle associated with the milli-weak
strength, and a large CP-violating phase associated with the gross
violation of the symmetry.

The mixing of three neutrino mass states, $m_1 < m_2 < m_3$, is described
by the mixing matrix, given in two alternative forms
\ba
\left( \begin{array}{c} \nu_e\\ \nu_\mu \\ \nu_\tau \end{array}\right) &=&
\left( \begin{array}{ccc} U_{e1} & U_{e2} & U_{e3} \\
                          U_{\mu 1} & U_{\mu 2} & U_{\mu 3}\\
			  U_{\tau 1} & U_{\tau 2} & U_{\tau 3} 
       \end{array} \right)\left(\begin{array}{c} \nu_1\\ \nu_2\\ \nu_3
          \end{array}\right)\nn\\
&=& \left( \begin{array}{ccc} c_{12} & s_{12} & 0 \\
                       -s_{12}c_{23} & c_{12}s_{23} & s_{23} \\
                       s_{12}s_{23} & -c_{12}s_{23} & c_{23} \\
     \end{array}\right) \left(\begin{array}{c} \nu_1\\ \nu_2\\ \nu_3
          \end{array}\right)
\ea
The second form is written in the standard form \cite{ref8}, in terms
of the sines (cosines) of the two mixing angles $s_{12}, s_{23}
(c_{12}, c_{23})$. The sine of the third mixing angle $s_{13}$ has
been set equal to zero. With this approximation, the CP-violating
phase factor $e^{-i\delta_{13}}$ drops out of this phenomenological
mixing matrix. The atmospheric data \cite{ref1,ref6} indicates that
$(m_3^2-m_2^2) \cong 2.5\times 10^{-3}\eV^2$. The mixing is essentially maximal
\cite{ref1,ref6}, which means that $s_{23} = \sin 45^\circ ( = c_{23})$.
The depletion of atmospheric $\nu_\mu$ via $\nu_\mu \to \nu_\tau$, is
proportional to $|2U_{\mu 3} U_{\tau 3}|^2 \cong 1$. Our ansatz for
the functional form of the effective mixing angle $\theta_{23}$ is
\ba
\theta_{23} \cong 45^\circ &=& \tan^{-1}\left(\mp \sqrt{\frac{m_\mu}{m_\tau}}\right)
+ \tan^{-1}\left(\sqrt{\frac{m_2}{m_3}}\right)\nn\\
& = & \mp 14^\circ +  \tan^{-1}\left(\sqrt{\frac{m_2}{m_3}}\right)\nn\\
&& \Rightarrow s_{23} = \sin\theta_{23} \cong 0.707
\ea
We choose the positive sign for the first term, so that $m_2 < m_3$.
Then eq.~(2) and $(m_3^2-m_2^2) \cong 2.5\times 10^{-3}\eV^2$ together, give
\be
m_3 \cong 5.37\times 10^{-2}\eV, \;\;\; m_2\cong 1.94\times 10^{-2}\eV
\ee
Eq.~(2) defines an effective, positive mixing angle for neutrino mass states
in terms of a difference between two mixing angles which are given by the
square roots of mass ratios. The larger (positive) angle is given by a
ratio of neutrino masses, and the smaller angle which can be negative or positive,
by a ratio of charged-lepton masses. The motivation for the square root
of mass ratio, is the physical idea that the mass $m_2$ is generated by a
second-order ``radiative correction'' proportional to $(g^2/m_3)$ which arises
from a coupling to the  larger mass $m_3$, with a coupling parameter
$g = (m_3\tan\theta)$ and $\tan\theta = \sqrt{m_2/m_3}$. The sign of such
a coupling parameter is a physically relevant variable; we take it as determining the
relative sign of the first term in eq.~(2) above. Following this argument, the
other effective mixing angle $\theta_{12}$ is
\ba
\theta_{12} &=& \tan^{-1}\left( \mp \sqrt{\frac{m_e}{m_\mu}}\right) +
\tan^{-1}\left( \sqrt{\frac{m_1}{m_2}}\right)\nn\\
&=& \mp 4^\circ + \tan^{-1}\left( \sqrt{\frac{m_1}{m_2}}\right)
\ea
If the sign of the first term were to be positive, we postulate that
the sum would be the same as in eq.~(2), i.~e.~$\theta_{12}\cong 45^\circ$.
This is done because of a corresponding similarity in the two analogous
equations in the quark sector, which are discussed below  (eq.~(8)). 
This determines $\sqrt{m_1/m_2}$, and hence
\be
m_1\cong 0.755 m_2 \cong 1.46\times 10^{-2}\eV \Rightarrow (m_2^2-m_1^2)
\cong 1.63\times 10^{-4}\eV^2
\ee
The ``hierarchal'' character\cite{ref6} of squared-mass differences
is thus calculated. With the  absolute value of both angles now fixed
on the right-hand side of eq.~(4), we take the negative sign of the
first term as determining the actual value of $\theta_{12}$
\be
\theta_{12} \cong -4^\circ+41^\circ = 37^\circ \Rightarrow \tan^2\theta_{12}\cong 0.56
\ee
This negative sign follows the pattern in the quark sector which gives rise
to one smaller mixing angle (the second equation in eq.~(8) below). The calculated
results in eqs.~(5,6) for $(m_2^2-m_1^2)$ and $\tan^2\theta_{12}=\tan^2\theta_\sol$ are well
within the allowed range, and are in fact, close to best-fit values from
the recent analyses \cite{ref4,ref5} of the KamLAND data (see fig.~1 and
table 1 in ref.~4, and fig.~1 in ref. 5). The mixing of atmospheric neutrinos
is controlled by $|2U_{\mu 3} U_{\tau 3}|^2 = \sin^2 2\theta_{23}\cong 1$, by
assumption. The disappearance of solar $\nu_e$ is controlled by the
calculated quantity
\ba
|U_{e1}U_{e2}| &=& \left| \left\{ |U_{e2}U_{\mu 2}|^2 + |U_{e2}U_{\tau 2}|^2\right\}^{1/2}\right|\nn\\
&=& |\cos\theta_{12}\sin\theta_{12}| \cong 0.48 < |\cos\theta_{23}\sin\theta_{23}| =
|U_{\mu 3}U_{\tau 3}| = 0.5
\ea
The last inequality reflects an important aspect of the data \cite{ref4,ref5}. The
solar neutrino mixing appears to be not maximal, whereas the atmospheric neutrino
mixing is essentially maximal, i.~e.~$\tan^2\theta_{12}\cong 0.56 < \tan^2\theta_{23}\cong 1$.
From the second form in eq.~(7) the solar $\nu_e$ disappear about equally
to $\nu_\mu$ and $\nu_\tau$. As an example of the sensitivity of the results
to small changes, $m_{1,2} \rightarrow m_{1,2}/\sqrt{2}$ gives $(m_1^2-m_2^2)
\cong 0.8\times 10^{-4}\eV^2$ with $\tan^2\theta_{12} \cong 0.56$ (these are about the
simultaneous best-fit numbers from the data analysis in ref.~4). With $m_3$
unchanged, $(m_3^2-m_2^2)$ and $\sin^2 2\theta_{23}$ change little, 
to $\sim 2.7\times 10^{-3}\eV^2$
and $\sim 0.98$, respectively.

Consider the equations in the quark sector \cite{ref9} which are the analogues
of eq.~(4) and eq.~(2), respectively (we denote the quark mixing angles by
$\tilde{\theta}_{12}$, $\tilde{\theta}_{23}$, and the sines by $\tilde{s}_{12}$,
$\tilde{s}_{23}$).
\ba
\tilde{\theta}_{12} &=& \tan^{-1}\left(\mp \sqrt{\frac{m_u}{m_c}}\right) +
\tan^{-1}\left(\sqrt{\frac{m_d}{m_s}}\right) \cong -1.8^\circ + 13.7^\circ \cong 11.9^\circ\nn\\
&\Rightarrow& \tilde{s}_{12} \cong 0.21\nn\\
\tilde{\theta}_{23} &=& \tan^{-1}\left(\mp \sqrt{\frac{m_c}{m_t}}\right) +
\tan^{-1}\left(\sqrt{\frac{m_s}{m_b}}\right) \cong -5^\circ + 8^\circ \cong 3^\circ\nn\\
&\Rightarrow& \tilde{s}_{23} \cong 0.05
\ea
We have taken the negative sign for the smallest angles. We use \cite{ref8} $m_u\cong 1.5\MeV$,
$m_d\cong 5\MeV$, $(m_s/m_d)\cong 17$, $m_c\cong 1.4\GeV$, $m_b\cong 4.5\GeV$, and $m_t\cong 175\GeV$. The
approximate nature of the numbers obtained from eqs.~(8) is also related to the
approximate nature of the definite quark masses used. An approximate relation
for the Cabibbo angle $\tilde{\theta}_{12}$, involving square roots of mass ratios, is long
known. \cite{ref10,ref11} Two additional interesting features of eqs.~(8) have been
commented upon \cite{ref9}: 
(1) the near cancellation which occurs for the effective $\tilde{\theta}_{23}$,
and (2) if the positive sign were to be chosen for both of the smallest angles,
then $\tilde{\theta}_{12} \sim \tilde{\theta}_{23}$. This is the motivation for
the postulate following eq.~(4) in the lepton sector, which determines $(m_1/m_2)$. (This common
angle being smaller than $\sim 45^\circ$ here, is of course related to the extreme
smallness of $m_{u,d}$ i.~e.~the approximate chiral SU(2) invariance of effective strong
interactions.)
With the minus sign, the near cancellation in determining the effective angle
$\tilde{\theta}_{23}$, suggests that this angle might be considered to be ``accidentally''
significantly smaller than $\tilde{\theta}_{12}$. Similarly, the non-maximal mixing
of the solar neutrinos need not appear as ``unnatural''. None of the above
considered effective mixing angles $\theta_{ij} (\tilde{\theta}_{ij})$ in the lepton
and the quark sectors need to be considered as intrinsically small. The exception
appears to be the empirical, non-zero $\tilde{s}_{13} < 0.005$ in the quark
sector \cite{ref8}. However, this very small number is associated with the phase
factor $e^{-i\delta_{13}}$ in the quark-mixing matrix element $|U_{u3}|$ with
empirically, $\delta_{13}>45^\circ$ \cite{ref8}. As remarked in the introduction, this
can suggest the existence of an effective \cite{ref12} milli-weak interaction
which grossly violates CP invariance, and which gives a dynamical basis for
the phenomenological parameterization of CP noninvariance which is contained
in the unitary KM matrix.

It is instructive to compare our above results for $m_1$, $m_2$, $m_3$ and
$|U_{\mu 3} U_{\tau 3}|$, $|U_{e1} U_{e2}|$, with those of another recent
calculation of the absolute values of three, hierarchal neutrino masses \cite{ref13},
which is based upon certain hypothetical, analogous structures for mixing in
the lepton and quark sectors. These results are \cite{ref13} $m_3 = (5.2-5.4)\times 10^{-2}\eV$,
$m_2 = (1.0-1.3) \times 10^{-2}\eV$, $m_1 = (0.2-0.5) \times 10^{-2}\eV$; and
$|U_{\mu 3} U_{\tau 3}| \cong (0.39-0.40)$, $|U_{e1}U_{e2}| \cong (0.42-0.44)$.
There are two essential differences to our results. In contrast to eq.~(7), here
$|U_{\mu 3}U_{\tau 3}| < |U_{e1}U_{e2}|$, which means that the mixing of solar
neutrinos is closer to maximal than the mixing of atmospheric neutrinos. This
does not appear to be the case, from analyses of the present experimental data \cite{ref4,ref5}.
The magnitude of $|U_{e1}U_{e2}|$ itself, is significantly smaller than our value;
this is related to a significantly smaller lowest mass $m_1$.

In conclusion, we have shown that the present data on the mixing of atmospheric
neutrinos and simple arguments based upon dynamical generation of hierarchal mass
structure, together with certain analogies to quark mixing, allow one to estimate the absolute values
of three neutrino masses, and also the non-maximal mixing relevant for
disappearance of KamLAND electron antineutrinos. There is a suggestion that the
mixing matrix (empirically, for quarks) embodies a dynamical origin for a large
CP-violating phase which is associated with an intrinsically, very small mixing angle.
A possible implication is the presence of additional loop diagrams in
$B$ decays, with dependence upon squared four-momentum transfer as probed by
a virtual gluon.
An experimental effect might be a failure
of the expected near equality \cite{ref14} of the effective mixing angle
for the mixing of $B^0$ states as determined in two different processes,
$B^0(\ol{B}^0)\to J/\psi + K_{\mathrm{S}},\; \phi+K_{\mathrm{S}}$.\cite{ref15}

\end{document}